\documentclass[twocolumn,showpacs,preprintnumbers,amsmath,amssymb,PRB]{revtex4}
\usepackage{graphicx}
\usepackage{dcolumn}
\usepackage{bm}

\begin{document}

\title{Current and strain-induced spin polarization in InGaN/GaN superlattices}

\author{H. J. Chang, T. W. Chen, J. W. Chen, W. C. Hong, W. C. Tsai, Y. F. Chen, and G. Y. Guo}
\address{Department of Physics, National Taiwan University, Taipei 106, Taiwan}

\begin{abstract}
The lateral current-induced spin polarization in InGaN/GaN superlattices (SLs) without an applied magnetic field is reported. The fact that the sign of the nonequilibrium spin changes as the current reverses and is opposite for the two edges provides a clear signature for the spin Hall effect. In addition, it is discovered that the spin Hall effect can be strongly manipulated by the internal strains. A theoretical work has also been developed to understand the observed strain induced spin polarization. Our result paves an alternative way for the generation of spin polarized current.
\end{abstract}


\maketitle

It has been proposed theoretically that a transverse spin current, the so called spin Hall current, can be generated in strongly spin-orbit coupling systems by external electric field~\cite{Dyakonov,Datta,Hirsch1,Edelstein,Ganichev,Murakami,Sinova}. This novel phenomenon offers the exciting possibility of pure electric driven spintronics in semiconductors, which can be more readily integrated with current industry. Recently, the observations of the spin Hall effect have been reported~\cite{Kato,Wunderlich,Sih} in GaAs based devices, which has generated an intensive research~\cite{Murakami3,Rashba2,Guo,Burkov,Hirsch,Sinitsyn}. However, up to now, most of the experimental measurements were concentrated on GaAs system. More clear evidences to confirm the spin Hall effect in other semiconductors are still needed~\cite{Stern}. Nitride semiconductors probably represent the most studied material system in semiconductor community for the last decade. InGaN/GaN superlattices have become particularly attractive because of their wide applications in light-emitting diodes (LEDs) ~\cite{Nakamura1} and high efficiency laser diodes (LDs) ~\cite{Nakamura3}. In this letter, we provide the first experimental observation of current-induced spin polarization in InGaN/GaN superlattices. In addition, we discover that the spin Hall effect can be strongly manipulated by the internal strains. In order to have a fundamental understanding of the strain-induced spin Hall effect, a theoretical investigation has also been performed. Because strains are commonly present in semiconductor multilayers and superlattices due to lattice mismatch, our result therefore provides a convenient way to control the magnitude of the spin Hall effect, which should be very useful for the future applications. Furthermore, this strain dependence enables to resolve the intensive debate about whether the intrinsic spin Hall effect remains valid beyond the ballistic transport regime. 

The n-type InGaN/GaN superlattices were grown by a horizontal flow metalorganic chemical vapor deposition on (0001) sapphire substrate. All structures consist of 30 nm buffer layer, followed by a 2 $\mu$m epitaxial GaN layer and then 50 periods of InGaN/GaN superlattices. Superlattices are composed of periods of alternating 30 $\AA$  GaN barriers and 50 $\AA$  In$_{x}$Ga$_{1-x}$N wells with x $\sim$  0.15. The superlattices was doped with Si, and the electron concentration is 2x$10^{12}$ $cm^{-2}$ according to the Hall measurement. For the cross sectional photoluminescence (PL) measurement, the sample was excited by a He-Cd laser working at 325 nm, dispersed by a SPEX 0.85 m double-grating spectrometer, and the spectra was detected by a photomultiplier tube. Figure 1 shows the experimental arrangement for measuring the current-induced spin polarization. The current flow was stabilized with a value of 80 mA. The laser light was incident along the c-axis, which in turn was normal to the growth surface, and the polarized PL spectra were collected along the edge direction of the sample. The cross sectional PL spectra with right-handed ($\sigma^{+}$) and left-handed ($\sigma^{-}$) modes were analyzed with a $\lambda$/4 wave plate and linear polarizer. For applying an electric field, ohmic contacts were formed by depositing indium drops and a constant current was applied by Keithley 236 source measure unit.
\begin{figure}
\begin{center}
\includegraphics[width=5cm]{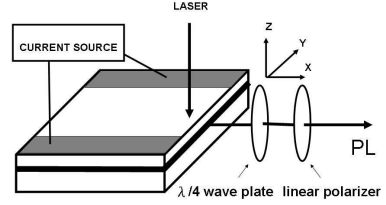}
\end{center}
\caption{The schematic diagram of the edge photoluminescence measurement with different angle of polarizer} 
\end{figure}

Figure 2(a) shows the cross sectional PL spectra with $\sigma^{+}$ and $\sigma^{-}$ polarizations of the InGaN/GaN superlattices without current flow. The main PL signal originates from the recombination of electron-hole pairs in the quantum wells, and the weak modulation on the spectrum is due to the interference effect~\cite{Lin}. The PL spectra were carefully checked by recording zero difference between $PL_{\sigma^{+}}$ and $PL_{\sigma^{-}}$ spectra when no current was delivered through the sample. Figures 2(b) shows the intensity variation of the cross sectional PL spectra at 13 K with different orientation of polarization when the current flow is turned on. Here, the angle of 0 degrees is defined as the polarization is parallel to the c-axis of the sample. As can be seen in Figs. 2(b), it reveals a dramatic change in the PL intensity with respect to the polarized angle when the current flow is turned on. With the positive current flow (along the y-direction), the PL signal of the $\sigma^{-}$ polarization is dominant as compared to that of the $\sigma^{+}$ polarization. On the other hand, the $\sigma^{+}$ polarization PL intensity is dominant for the negative current flow (along the -y-direction), i.e., the PL polarization is inverted. 
\begin{figure}
\begin{center}
\includegraphics[width=6cm]{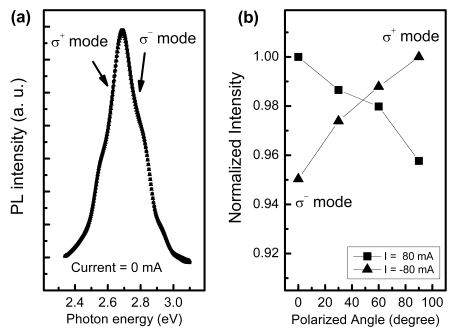}
\end{center}
\caption{(a) Cross sectional photoluminescence spectra with $\sigma^{+}$ and $\sigma^{-}$ polarizations of the InGaN/GaN superlattices in absence of a current. (b) shows the intensity variation of the cross sectional PL spectra with different orientation of polarization when the current flow ( I = +/- 80 mA ) is turned on at 13 K.} 
\end{figure}

To illustrate the effect of the current flow on the PL spectra in more detail, the difference of the PL intensity between $\sigma^{+}$ and $\sigma^{-}$ polarizations is shown in Fig. 3. The base line (dotted) is taken with the current turned off. The result seems containing a large fluctuations, which may arise from the effect of scattering due to a large defect density existing in nitride semiconductors~\cite{Kato}. Even with the large fluctuations, it reveals an important feature that the difference of the edge PL spectra inverses its sign as the electric current is reversed. In other words, the PL polarization is inverted as the direction of the current flow is reversed. This is a direct evidence of the current-induced spin orientation, in complete agreement with the results on the spin photocurrents in previous reports~\cite{Kato,Wunderlich,Sih}. It is also worth noting that we did not detect a measurable anisotropy for the in-plane PL spectra with the current flow, and there is also no measurable anisotropy in the edge PL spectrum for the band to band emission when the current is turned off. Thus, the optical anisotropy is a unique property related to the current-induced spin polarization. In addition, to further confirm that the detected optical anisotropy is indeed due to nonequilibrium spin accumulation at sample edges, we have performed and found that the PL polarization is opposite for the two edges.
\begin{figure}
\begin{center}
\includegraphics[width=6cm]{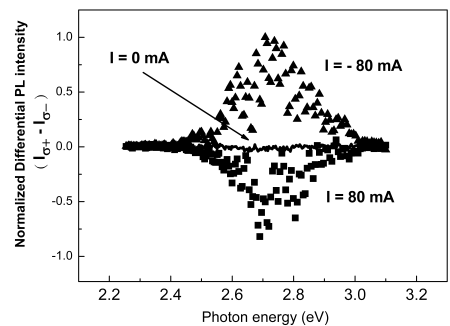}
\end{center}
\caption{Differential spectra of polarized photoluminescence. Base line is taken with the current turned off.} 
\end{figure}

Let us now briefly outline the underlying mechanism that the nonzero differential edge polarized PL spectra detected in the absence of an external magnetic field is indeed an experimental demonstration of the spin-polarization effect caused by asymmetric band occupation due to the current flow~\cite{Mal'shukov}. When a $\sigma$-spin electron in the $\mathbf{k}$ state of the conduction band makes a radiative transition to the topmost heavy-hole subband, the probability of this process depends on $\sigma$ because the population of hole at the $\mathbf{k}$ state in the split up subband may be different from that in the split down subband. Hence, at $\mathbf{k}$ the conduction-band electron spin will be polarized along the $\sigma$ direction if an electron with ($\sigma$) spin has a higher recombination rate. However, since the hole spin orientation at -$\mathbf{k}$ is reversed with respect to that at $\mathbf{k}$, the spin polarization of the conduction band electron at the -$\mathbf{k}$ state will be along the -$\sigma$ direction. Consequently, if the momentum distributions of both the electron gas and the hole gas are isotropic, there will be no net spin polarization of the conduction-band electrons. On the other hand, if such momentum distributions are anisotropic, the generation of spin polarization becomes possible.

To further elucidate the unique behaviors of the spin Hall effect in the InGaN/GaN superlattices, we have measured the edge polarized PL spectra under different excitation power as shown in Fig. 4. We can clearly see that the degree of circular polarization P decreases with increasing pumping power, where P is defined as
\begin{equation}
   P=\frac{I_{\sigma+}-I_{\sigma-}}{I_{\sigma+}+I_{\sigma-}},
    \label{TH3}
\end{equation}
This interesting result is quite subtle. To understand this intriguing behavior, let us recall the reported photoelastic effect in InGaN/GaN multiple quantum wells~\cite{Lin}. It has been shown that increasing excitation intensity will enhance the screening of piezoelectric (PZ) field due to spatial separation of photoexcited electrons and holes. At the first glance, it seems clear that our observation can be simply explained in terms of the reduced Rashba effect due to the screened electric field. However, one has to keep in mind that nitride semiconductors are excellent piezoelectric materials. Through the converse PZ effect, a change in the built-in electric field can induce a sizable change in the strain. According to the previous study~\cite{Guo,Gvozdic}, a small change of strain can have a significant influence on the Rashba effect, which has been considered as one of the main causes for the spin Hall effect~\cite{Dyakonov,Datta,Hirsch1,Edelstein,Ganichev,Murakami,Sinova}. Therefore, to complete our understanding, it will be worthwhile to take the strain effect into account.
In order to demonstrate that the pumping laser can indeed change the internal strain in the studied sample, we have measured the Raman scattering of the InGaN/GaN superlattices under different excitation density as shown in Fig. 5. It clearly indicates that the phonon frequency decreases with increasing pumping power, which implies that the internal strain of the heterostructure due to lattice mismatch can be changed by external excitation. The redshift in Raman spectra is unlikely due to local heating of the sample, because the corresponding PL peak has a blueshift, instead of a redshift (not shown here). If local heating is significant, the band gap will be shrinked, which will result in a redshift of the PL spectra.  In addition, the fact that the PL peak of GaN layer remains unchanged further supports a negligible heating effect. The strain $\epsilon$ change can be calculated according to phonon shift $\Delta$$\omega$ by the expression~\cite{Kontos}
\begin{equation}
   {\epsilon}=\frac{{{\Delta}\omega}}{2a-2b\frac{C_{13}}{C_{33}}},
    \label{TH2}
\end{equation}
where $a$ and $b$ are phonon deformation potentials ,and $C_{13}$ and $C_{33}$ are the elastic constants, respectively. The calculated strain is shown in the inset of Fig. 5, where the deformation potentials a, b and the elastic constants $C_{13}$, $C_{33}$ were estimated by the extrapolation of the data for GaN and InN~\cite{Kontos}.
\begin{figure}
\begin{center}
\includegraphics[width=4.5cm]{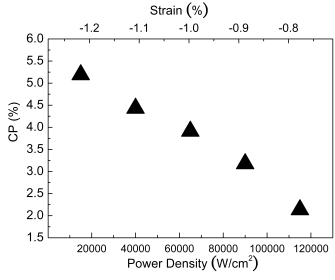}
\end{center}
\caption{The degree of circular polarization as a function of excitation density and corresponding strain. The strain is calculated by the Raman spectra in Fig. 5.The in-plane electric field is applied in -y direction.} 
\end{figure}
\begin{figure}
\begin{center}
\includegraphics[width=6cm]{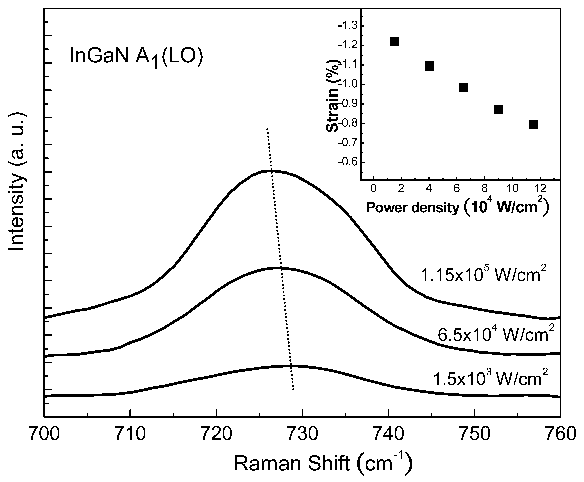}
\end{center}
\caption{Raman shift as a function of excitation density. In the inset, the calculated strain as a function of excitation density. It shows that the internal strain in InGaN/GaN superlattices can be manipulated by external excitation.} 
\end{figure}
\begin{figure}
\begin{center}
\includegraphics[width=6cm]{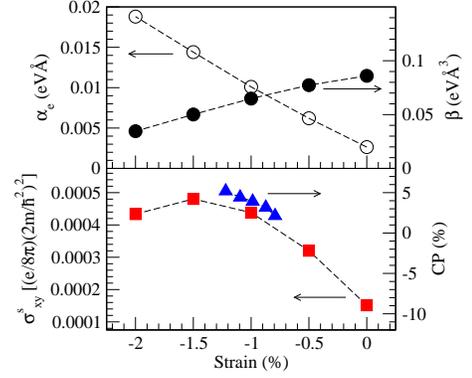}
\end{center}
\caption{(a) The theoretical effective Rashba
coupling $\alpha_e=\lambda+\alpha$ ($eV\AA$) (open circle) and BIA coupling
$\beta$ ($eV\AA^3$) (solid circle) vs. the in-plane
strain. (b) the theoretical spin Hall
conductivity ($\sigma^{s}_{xy}$) (solid square) and the degree of circular polarization (CP) (solid triangle)
vs. the in-plane strain. The in-plane electric field is applied in -y direction. } 
\end{figure}

To establish the fundamental background of the strain-dependent spin Hall effect in 
InGaN/GaN superlattices, a theoretical investigation of the
spin Hall current in n-type two-dimensional (2D)
wurtzite structure is performed.
Taking into account of the strong polarization field at the
interface alone the $c$-axis of wurtzite structure and
the bulk inversion asymmetry (BIA), the 2D
Hamiltonian can be written as~\cite{Zor96}
\begin{equation}
H =\frac{\hbar k^{2}}{2m}+(\alpha_{e}+\beta
k^{2})(\sigma_{x}k_y-\sigma_yk_x)\\,
\label{WBIA2}
\end{equation}
where $\alpha_{e}=\lambda+\alpha$ is the effective Rashba
coupling. The coupling $\lambda$ is induced by the built-in
electric field due to the piezoelectric effect.  The coupling
$\alpha$ and $\beta$ are due to the BIA effect. To estimate the
coupling parameters $\alpha_{e}$ and $\beta$, we have performed
{\it ab initio} relativistic band structure calculations for 
GaN under several compressive in-plane strains (0 $\sim$ -2.0 \%) 
using the highly accurate all-electron full-potential linearized augmented
plane wave (FLAPW) method~\cite{wien2k}. The details of the
calculations will be reported elsewhere~\cite{guo06}. 
By fitting the 2D Hamiltonian (Eq. 3) to the calculated 
spin-orbit coupling splitting of the lowest conduction bands 
near $k = 0$ as a function of the wavevector $k_x$, we obtain 
the theoretical $\alpha_{e}$ and $\beta$, as shown in Fig. 6(a) as
a function of the in-plane strain.
We note that the theoretical $\alpha_{e}$ values are in the 
same order of magnitude of the effective Rashba coupling
measured on AlGaN/GaN~\cite{Cho}. 
Since the magnitude of the in-plane strain in InGaN/GaN is 
close to that in AlGaN/GaN, the magnitude of built-in electric field 
is found to be in the same order of magnitude in both systems
($\sim 1MV/cm$)~\cite{Gran,Gran2}. Therefore, we may expect
InGaN/GaN and AlGaN/GaN to have similar coupling parameters $\alpha_{e}$ 
and $\beta$, and hence the theoretical $\alpha_{e}$ and $\beta$
(Fig. 6a) will be used in the following spin Hall conductivity
calculations.

Within the linear-response Kubo formalism (see, e.g., ~\cite{chen06}), we find that the spin Hall conductivity for the 2D Hamiltonian of Eq. (3) 
is given by
\begin{equation}
\sigma^{s0}_{xy}=-\sigma^{s0}_{yx}
\approx\frac{e}{8\pi}[1+\frac{2}{3}\alpha_{e}\beta(\frac{2m}{\hbar^2})^{2}],
\label{sc2}
\end{equation}
under the condition that
$\alpha_{e}$ $(\sim 10^{-2} eV\AA)$ is larger than $\beta k^2_{\pm}$ 
$( 10^{-3}\sim 10^{-4} eV\AA)$, where the $k_{+}$ and $k_{-}$ are
the Fermi momenta of the two spin-split bands, respectively. The spin Hall conductivity would be a universal constant, as found before~\cite{Sinova},
if we set $\beta=0$. However, it is well known~\cite{Jai} that in the presence of
scattering due to, e.g., the disorder, 
the spin Hall conductivity due to the $k$-linear
coupling term would vanish. 
Therefore, the observed spin Hall effect
would be predominantly due to the $k$-cubic term, i.e.,
$\sigma^{s}_{xy}\approx\frac{e}{8\pi}\frac{2}{3}(\frac{2m}{\hbar^2})^2\alpha_e\beta$.
This clearly shows that the spin Hall effect in the wurtzite semiconductor
superlattices and multilayers could be enhanced by manipulating the
spin-orbit coupling parameters $\alpha_e$ and $\beta$ which in term
can be tuned by the compressive in-plane strain via the piezoelectric
field, as demonstrated by
our {\it ab initio} relativistic band structure calculations (Fig. 6a).
The monotonic increase of the calculated spin Hall conductivity
as a function of the compressive in-plane strain is in good qualitative agreement with 
our experimental observation as shown in Fig. 6(b). This correlation implies that the 
circularly polarized PL spectra observed above are indeed due to the spin accumulation, 
which can be fine tuned by the internal strains. We therefore open a new possibility 
to manipulate the spin Hall effect in semiconductor heterostructures.

In conclusion, we have investigated the current-induced spin polarization in InGaN/GaN superlattices (SLs). It is found that the degree of polarization changes sign as the direction of the current flow is reversed. In addition, we also found that the internal strain in InGaN/GaN superlattices can be used to manipulate the magnitude of the spin Hall effect. The corresponding theoretical calculation has also been developed to interpret the strain dependence of the spin Hall effect. The strain-induced spin Hall effect discovered here paves an alternative way for the creation of spin polarized current, which should be useful for the realization of the future applications in spintronics. In view of the wide range of applications in nitride semiconductors, our study shown here should be very useful and timely.

This work was supported by the National Science Council and Ministry of Education of the Republic of China.

\end{document}